\documentclass[final,twocolumn,3p,times]{elsarticle}
\usepackage{amssymb}
\usepackage{amsmath}
\usepackage{subcaption}
\usepackage{physics}
\usepackage{orcidlink}
\journal{Physics Letters A}

\begin{document}

\begin{frontmatter}

%% Title, authors and addresses

%% use the tnoteref command within \title for footnotes;
%% use the tnotetext command for theassociated footnote;
%% use the fnref command within \author or \affiliation for footnotes;
%% use the fntext command for theassociated footnote;
%% use the corref command within \author for corresponding author footnotes;
%% use the cortext command for theassociated footnote;
%% use the ead command for the email address,
%% and the form \ead[url] for the home page:
%% \title{Title\tnoteref{label1}}
%% \tnotetext[label1]{}
%% \author{Name\corref{cor1}\fnref{label2}}
%% \ead{email address}
%% \ead[url]{home page}
%% \fntext[label2]{}
%% \cortext[cor1]{}
%% \affiliation{organization={},
%%            addressline={}, 
%%            city={},
%%            postcode={}, 
%%            state={},
%%            country={}}
%% \fntext[label3]{}

\title{Fluctuation-stable generalized entropy probes of spectral heterogeneity}

\author{Arpita Goswami\,\orcidlink{0009-0003-2241-7034}}
\affiliation{organization={Department of Physics, Indian Institute of Technology Tirupati},%Department and Organization
             addressline={Yerpedu}, 
%             city={Tirupati},
            postcode={517619}, 
%            state={Andhra Pradesh},
            country={India}}
 %This line break forced with \textbackslash\textbackslash

\begin{abstract}
Generalized entropy measures are widely used to characterize localization and multifractality, and the regime \(q>1\) is often empirically found to exhibit improved numerical stability and cleaner scaling behavior. Here, we develop a fluctuation-stability framework for generalized entropy diagnostics and show that weak-amplitude spectral fluctuations are amplified for \(q<1\) and suppressed for \(q>1\), thereby providing a theoretical basis for the physically robust \(q>1\) regime. A thermodynamic scaling analysis further identifies an asymptotically stable regime beyond a critical threshold. As an application, we introduce the entropy-gradient susceptibility \(\chi_q\) as a coarse-grained probe of spectral heterogeneity. Using the Aubry-Andr\'e and generalized Aubry-Andr\'e models, we demonstrate that \(\chi_q\) sharply distinguishes homogeneous localization transitions from mobility-edge coexistence regimes. Our results establish fluctuation stability as a guiding principle for generalized entropy diagnostics in quasiperiodic systems.

\end{abstract}

\end{frontmatter}

\section{Introduction}
In low-dimensional quantum systems, localization arises from the suppression of wave transport due to interference effects. While uncorrelated disorder localizes all single-particle states in one dimension, quasiperiodic lattices~\cite{QP_1, QP_2, QP_3, QP_4, QP_5, QP_6, QP_7, QP_8, QP_9, QP_10, QP_11} realize localization through deterministic spatial modulation and therefore provide a distinct route to nonergodic behavior. The Aubry-Andr\'e (AA) model constitutes the canonical example of quasiperiodic localization, exhibiting a self-dual transition in which the entire spectrum changes collectively from extended to localized behavior. More generally, however, many quasiperiodic systems host mobility edges (MEs)~\cite{ME_1, ME_2, ME_3, ME_4, ME_5, ME_6, ext1}, where localized and extended eigenstates coexist within different energy sectors of the same spectrum. This spectral coexistence plays an important role in transport, thermalization, and nonequilibrium dynamics across a broad class of quantum systems.

Generalized entropy measures and participation ratios are widely used to characterize localization, multifractality, and spectral structure in such systems. In practice, the regime \(q>1\) is often empirically preferred because it exhibits reduced fluctuations and cleaner scaling behavior. However, a systematic theoretical understanding of why the \(q>1\) sector behaves more robustly has remained largely unclear. At the same time, conventional localization diagnostics such as the inverse participation ratio (IPR)~\cite{ipr_1,ipr_2,ipr_3}, multifractal observables, and Lyapunov exponents~\cite{LE_1, LE_2} primarily characterize individual eigenstates and do not directly address how wavefunction structure reorganizes collectively across the spectrum. Entropy-based probes are increasingly employed across diverse disordered systems, from quasiperiodic lattices to correlated random potentials, underscoring the need for a general stability criterion.

In this work, we develop a fluctuation-stability theory for generalized entropy diagnostics in quasiperiodic systems using Tsallis entropy~\cite{Tsallis1988}. We show analytically that weak-amplitude spectral fluctuations are amplified for \(q<1\) and suppressed for \(q>1\), thereby providing a theoretical foundation for the physically robust \(q>1\) sector. A thermodynamic scaling analysis further reveals an asymptotically stable regime for \(q>3/2\), in which entropy fluctuations are progressively suppressed as the system size increases. Generalized participation ratios and entropy-based multifractal diagnostics are frequently analyzed in the regime \(q>1\) because this sector empirically exhibits improved numerical stability and smoother scaling behavior. We also verified the enhanced entropy fluctuations numerically in the $q<1$ region using a GAA model.

Motivated by these stability properties, we also introduce the entropy-gradient susceptibility as an energy-resolved response function that probes spectral restructuring across the quasiperiodic spectrum. Unlike conventional state-resolved localization measures, the susceptibility captures coarse-grained spectral heterogeneity through the energy variation of the generalized entropy profile. Within the fluctuation-stable sector \(q>1\), the susceptibility exhibits sharp, robust signatures associated with the coexistence of localized and extended states.
In quasiperiodic systems, atypical eigenstates can exert a disproportionate influence on spectral and transport properties due to their enhanced statistical weight and critical spatial structure. Conventional averaged diagnostics may retain such localized spectral irregularities. Recent cold-atom and photonic experiments have already observed mobility edges, making entropy-gradient susceptibility a timely diagnostic with direct experimental relevance. Motivated by this, we introduce an entropy-gradient susceptibility designed to resolve rapid eigenstate structural variations across the spectrum and thereby sensitively probe mobility-edge physics and rare-state-induced fluctuations. In the AA model, where the localization transition is spectrally homogeneous, the entropy profile evolves smoothly, and the susceptibility response remains weak. By contrast, the generalized Aubry-Andr\'e model exhibits pronounced and size-stable susceptibility peaks within the mobility-edge regime, reflecting rapid spectral restructuring induced by the coexistence of localized and extended eigenstates. These results establish fluctuation stability as a guiding principle for generalized entropy diagnostics of localization phenomena in quasiperiodic systems.

The remainder of the paper is organized as follows. Section~\ref{sec_1} introduces the models considered. Section~\ref{sec_2} presents the entropy-based formalism and defines the susceptibility. Section~\ref{sec_3} discusses the fluctuation stability analysis. Numerical results are presented in Section~\ref{sec_4}, followed by analytical insights in Section~\ref{sec_5}. Finally, Section~\ref{sec_6} summarizes the conclusions.

\begin{figure*}[t]
    \centering
    \begin{subfigure}[t]{0.45\textwidth}
        \includegraphics[width=\textwidth]{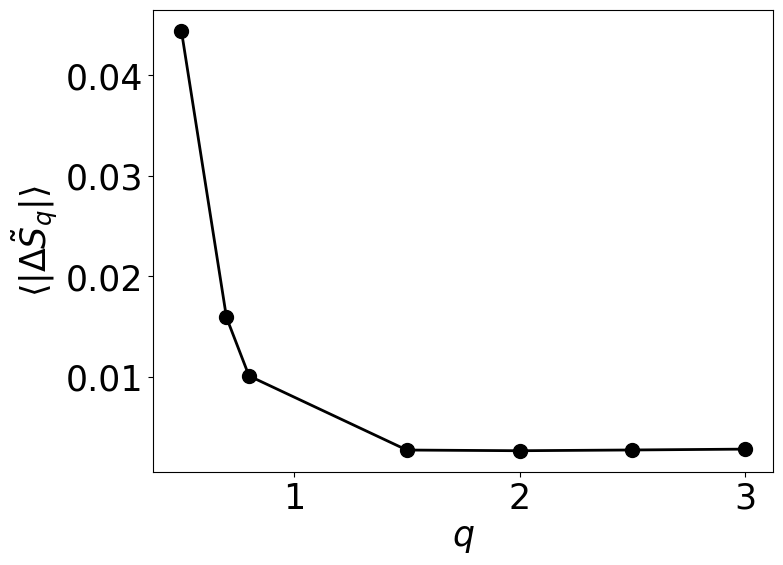}
        \caption{\( \langle|\Delta \tilde S_q|\rangle\) vs. \( q \)}
    \end{subfigure}
    \hspace{0.05\linewidth}
    \begin{subfigure}[t]{0.45\textwidth}
        \includegraphics[width=\textwidth]{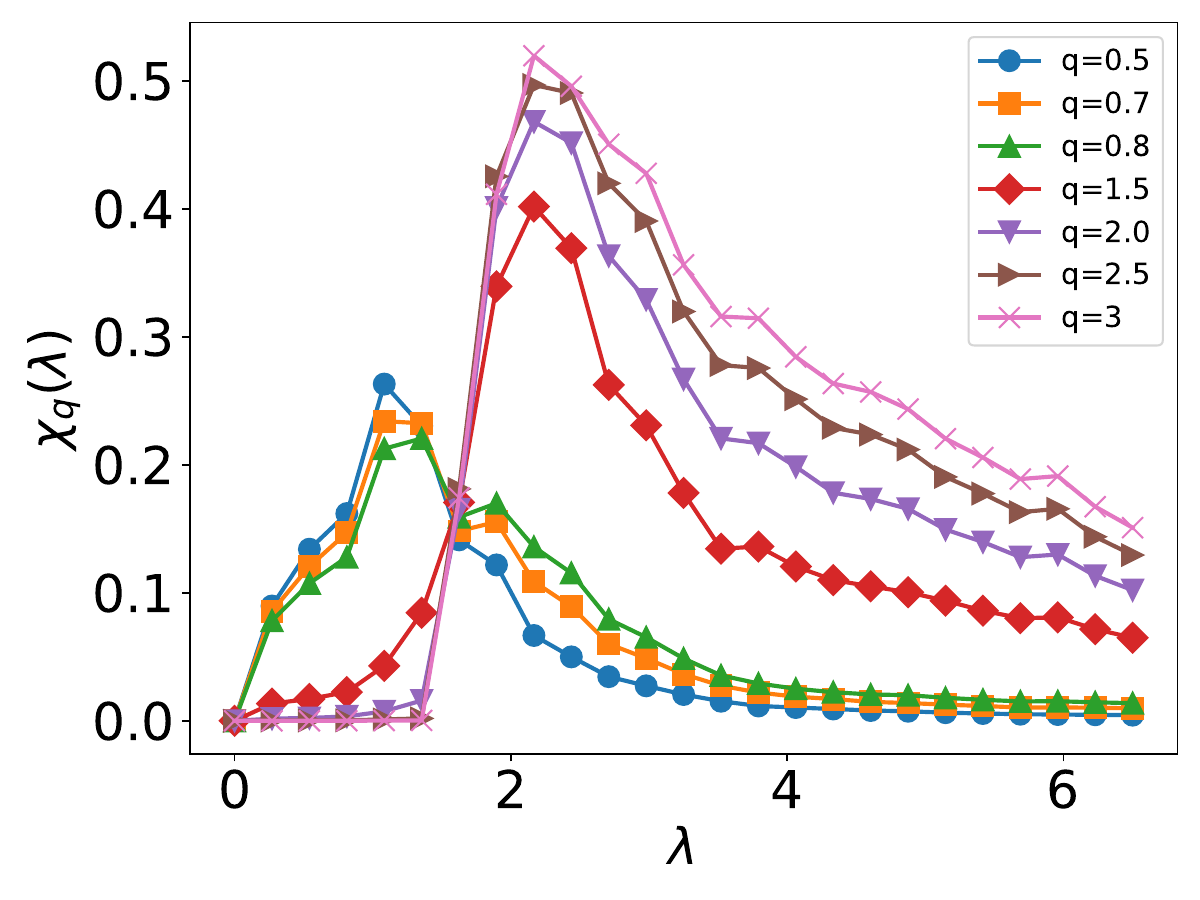}
        \caption{\( \chi_q(\lambda) \) vs. \( \lambda \).}
    \end{subfigure}
    \caption{(a) Average deviation of the normalized Tsallis entropy under small perturbations of the wavefunction amplitudes as a function of the entropic index q. The enhanced sensitivity for $q<1$ reflects the increased contribution of low-probability components, while the stable behavior for $q>1$ demonstrates the robustness of the entropy-based diagnostic. b) Entropy-gradient susceptibility $\chi_q(\lambda)$ of the GAA model as a function of the quasiperiodic potential strength $\lambda$ for $q=0.5, 0.7, 0.8, 1.5,2,2.5,3$. Both the calculations are performed for $L=600$.}
    \label{fig:fig1}
\end{figure*}

\begin{figure*}[htbp]
    \centering
    \begin{subfigure}[t]{0.43\textwidth}
        \includegraphics[width=\textwidth]{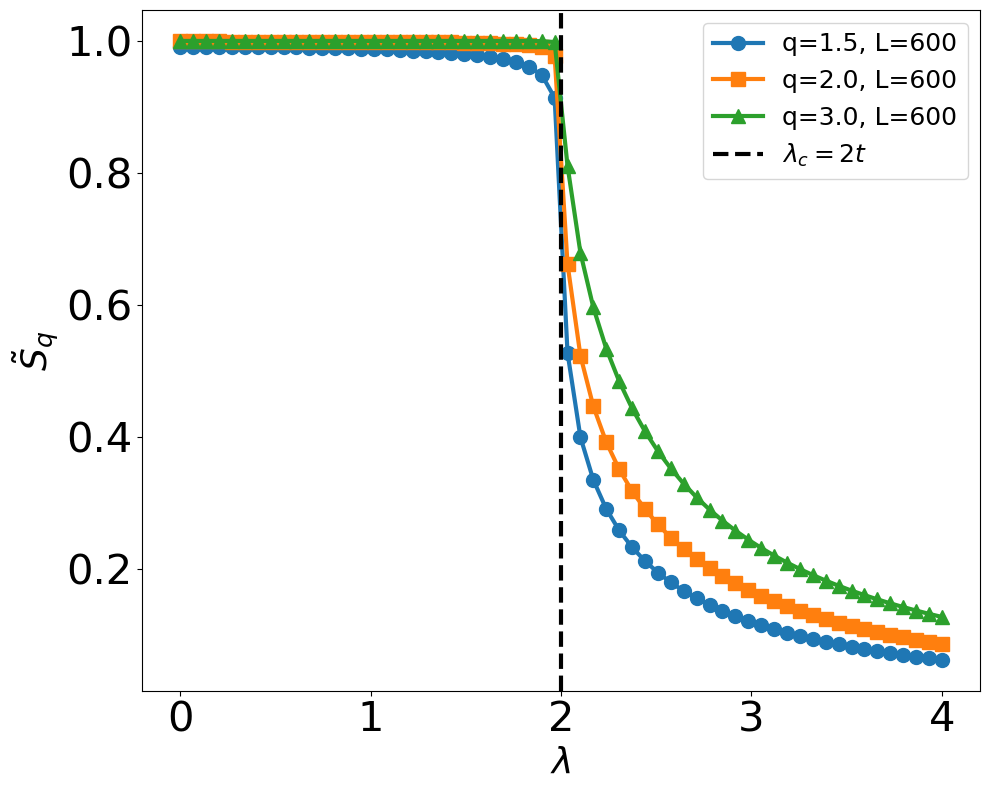}
        \caption{\( \tilde S_q\) vs. \( \lambda \) for \( q > 1 \)}
    \end{subfigure}
    \hspace{0.05\linewidth}
    \begin{subfigure}[t]{0.45\textwidth}
        \includegraphics[width=\textwidth]{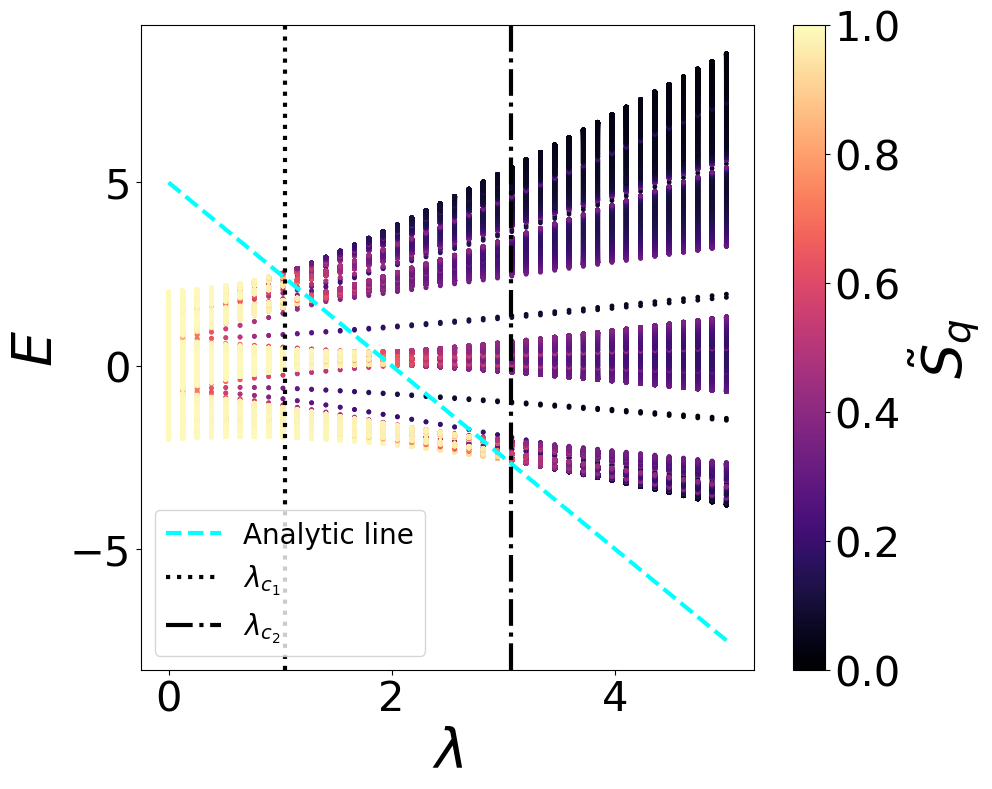}
        \caption{Contour plot of \( \tilde S_q \)}
    \end{subfigure}
    \caption{(a) Normalized Tsallis entropy $\tilde{S_q}$ as a function of quasiperiodic potential strength $\lambda$ for the Aubry-Andr\'e (AA) model at different entropic orders $q$, with system size $L=600$. (b) Normalized Tsallis entropy $\tilde S_q$ of the GAA model as a function of quasiperiodic potential strength $\lambda$ at $q=1.5$ and $L=600$.
    The entropy behavior shows the presence of MEs. }
    \label{fig:fig2}
\end{figure*}

\begin{figure*}[t]
    \centering
    \begin{subfigure}[t]{0.45\textwidth}
        \includegraphics[width=\textwidth]{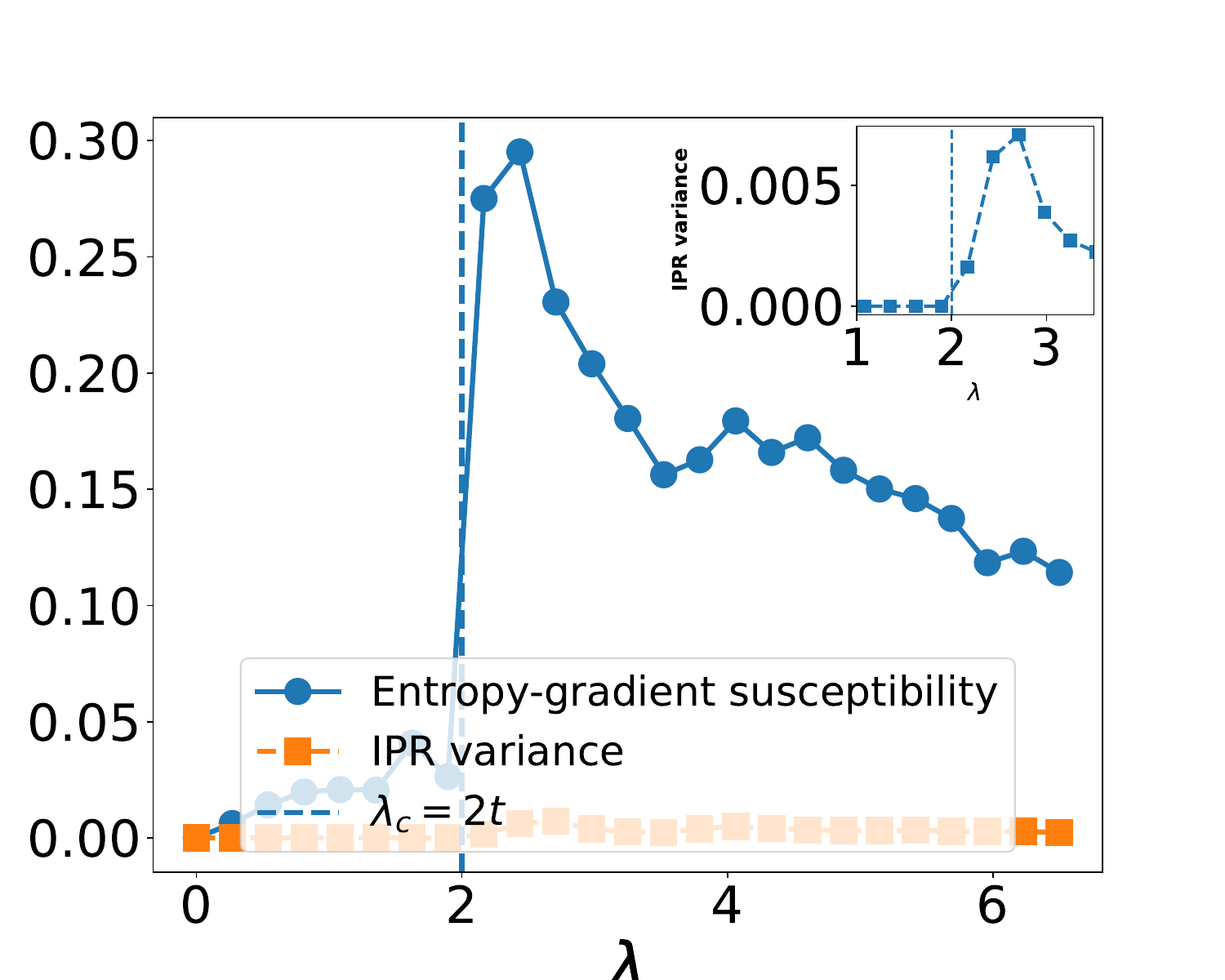}
        \caption{Plot for AA model}
    \end{subfigure}
    \hspace{0.05\linewidth}
    \begin{subfigure}[t]{0.45\textwidth}
        \includegraphics[width=\textwidth]{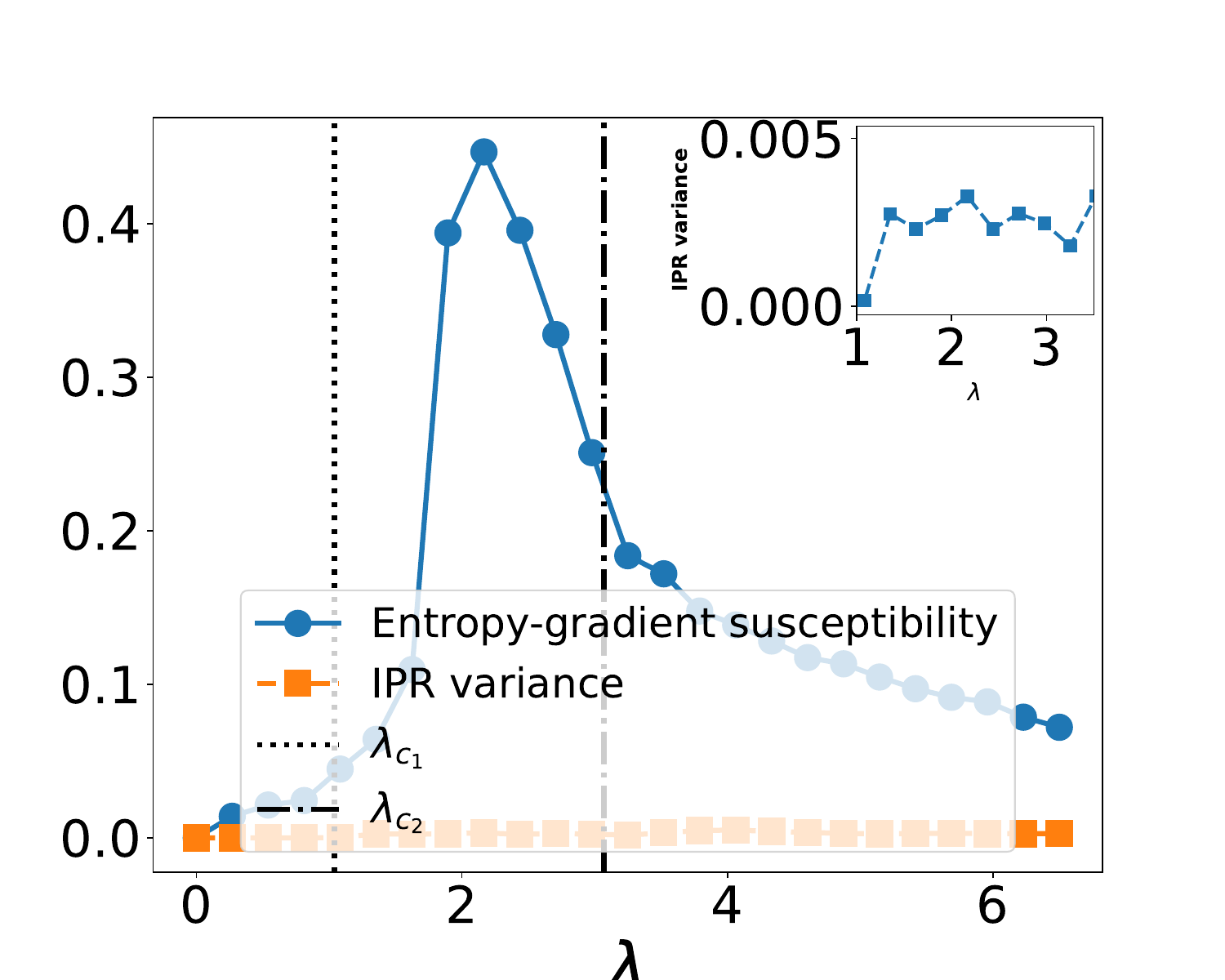}
        \caption{Plot for GAA model}
    \end{subfigure}
    \caption{(a)  Entropy-gradient susceptibility $\chi_q(\lambda)$ of the AA model as a function of $\lambda$. The absence of a pronounced peak reflects the lack of energy-resolved localization, consistent with a global localization transition. (b) Entropy-gradient susceptibility $\chi_q(\lambda)$ of the GAA model as a function of $\lambda$. A pronounced and systematically sharpening peak emerges at intermediate quasiperiodic strength, signaling enhanced spectral heterogeneity due to the coexistence of localized and extended eigenstates. Inset: IPR variance vs. $\lambda$.}
    \label{fig:fig3}
\end{figure*}

\section{Models}
\label{sec_1}

We consider two one-dimensional quasiperiodic lattice models that exhibit distinct localization scenarios: one without a mobility edge and one with mobility edges.

\subsection{Aubry-Andr\'e (AA) model}

The AA model is defined by
\begin{equation} \label{H_AA}
H_{\mathrm{AA}} = -t \sum_{n} \left( c_{n+1}^\dagger c_n + \text{h.c.} \right)
+ \lambda \sum_n \cos(2\pi \beta n + \phi)\, c_n^\dagger c_n ,
\end{equation}
  where $t$ is the nearest-neighbor hopping amplitude, $\lambda$ is the quasiperiodic potential strength, $\beta = (\sqrt{5}-1)/2$ is irrational number, and $\phi$ is a phase offset. Throughout, we set $t=1$.

The AA model is self-dual and undergoes a sharp localization transition at $\lambda_c = 2t$. For $\lambda < \lambda_c$, all eigenstates are extended, whereas for $\lambda > \lambda_c$, all eigenstates are exponentially localized. Since the entire spectrum localizes simultaneously, the AA model serves as a benchmark for a spectrally homogeneous transition.

\subsection{Generalized Aubry-Andr\'e (GAA) model}

The GAA model extends the AA model by introducing an energy-dependent quasiperiodic potential,
\begin{align} \label{H_GAA}
H_{\mathrm{GAA}} =& -t \sum_{n} \left( c_{n+1}^\dagger c_n + \text{h.c.} \right) \notag 
\\ & +  \sum_n
\frac{\lambda \cos(2\pi \beta n + \phi)}
{1 - \alpha \cos(2\pi \beta n + \phi)}
c_n^\dagger c_n ,
\end{align}
where $\alpha \in (0,1]$ controls the deviation from the AA limit. We fix $\alpha=0.4$ in the numerical calculations.

For $\alpha \neq 0$, the model exhibits an exact mobility edge determined by
\begin{equation} \label{ME_GAA}
\alpha E = 2t - \lambda.
\end{equation}
This closed-form relation provides a direct benchmark for testing the entropy-gradient susceptibility against the analytically known localization boundary.

\section{Tsallis entropy and entropy-gradient susceptibility}
\label{sec_2}

We introduce an entropy-based framework to characterize the energy-dependent structure of single-particle eigenstates. For a normalized eigenstate $\psi^{(n)}$ defined on $N$ lattice sites, the probability distribution is
\begin{equation}
p_i^{(n)} = |\psi_i^{(n)}|^2.
\end{equation}
The corresponding Tsallis entropy is defined as
\begin{equation}
S_q^{(n)} =
k_B\frac{1-\sum_i \left(p_i^{(n)}\right)^q}{q-1},
\end{equation}
which reduces to the Boltzmann-Gibbs-Shannon entropy in the limit $q\to1$. Since $S_q$ is directly related to generalized participation ratios, it provides a flexible characterization of wavefunction structure across different amplitude scales. We set $k_B =1.$

To compare different system sizes, we define the normalized entropy
\begin{equation}
\tilde S_q^{(n)}
=
\frac{S_q^{(n)}}{S_q^{\mathrm{max}}},
\qquad
S_q^{\mathrm{max}}
=
\frac{1-N^{1-q}}{q-1},
\end{equation}
where $S_q^{\mathrm{max}}$ corresponds to a completely extended state. Consequently, extended states satisfy $\tilde S_q^{(n)}\approx1$, while localized states yield smaller values.

The central quantity of this work is the entropy-gradient susceptibility,
\begin{equation}
\label{susceptibility}
\chi_q
=
\left\langle
\left|
\frac{d\tilde S_q(E)}{dE}
\right|
\right\rangle_E.
\end{equation}
Here, $\langle \dots \rangle_E$ represents an average over the entire energy spectrum, allowing $\chi_q$ to quantify the global spectral heterogeneity for a given potential strength $\lambda$. In systems with homogeneous localization behavior, the entropy profile varies smoothly with energy, resulting in a weak susceptibility response. In contrast, systems hosting mobility edges exhibit rapid spectral restructuring due to the coexistence of localized and extended states, producing pronounced peaks in $\chi_q$. The entropic index $q$ controls the relative sensitivity to dominant and low-probability amplitudes. As shown below, the regime $q>1$ yields stable and physically informative behavior, while $q<1$ enhances low-probability fluctuations and broadens the susceptibility response.

In practice, $\tilde S_q(E)$ is evaluated on a coarse-grained energy grid, and derivatives are computed after mild smoothing. We verified that the qualitative features of $\chi_q$, including peak structure and position, remain robust against variations in numerical procedures. It is worth mentioning that although the present analysis employs the Tsallis entropy, the framework can be generalized to other entropy measures with tunable amplitude sensitivity.

\section{Fluctuation stability of the Tsallis entropy framework} \label{sec_3}

A central question underlying the entropy-based characterization introduced in this work is the stability of the generalized entropy under weak perturbations of the wavefunction amplitudes. Numerical results presented in Fig.~\ref{fig:fig1}(a) show that the normalized Tsallis entropy exhibits markedly different fluctuation behavior across the entropic index $q$: the regime $q<1$ displays enhanced sensitivity and broad fluctuations, whereas the regime $q>1$ remains comparatively stable and robust. In this section, we provide an analytical explanation for this behavior and derive an upper bound for the entropy fluctuations.

\subsection{Perturbative response of the Tsallis entropy}

For a normalized single-particle eigenstate,
\begin{equation}
\sum_i p_i = 1,
\end{equation}
the Tsallis entropy is defined as
\begin{equation}
S_q = \frac{1-\sum_i p_i^q}{q-1}.
\label{eq:Tsallis}
\end{equation}

We now consider weak perturbations of the probability distribution,
\begin{equation}
p_i \rightarrow p_i + \delta p_i,
\end{equation}
subject to the normalization constraint
\begin{equation}
\sum_i \delta p_i = 0.
\end{equation}

Expanding the perturbed probability amplitudes to second order gives
\begin{equation}
(p_i+\delta p_i)^q
=
p_i^q
+
q p_i^{q-1}\delta p_i
+
\frac{q(q-1)}{2}p_i^{q-2}(\delta p_i)^2
+\mathcal{O}((\delta p_i)^3).
\end{equation}

Substituting into Eq.~(\ref{eq:Tsallis}) yields
\begin{align}
\delta S_q
&=
-\frac{q}{q-1}\sum_i p_i^{q-1}\delta p_i
\nonumber\\
&\quad
-\frac{q}{2}\sum_i p_i^{q-2}(\delta p_i)^2
+\mathcal{O}((\delta p_i)^3).
\label{eq:dSq}
\end{align}

Equation~(\ref{eq:dSq}) directly reveals the origin of the stability crossover across the entropic index $q$. The fluctuation response is weighted by powers of the probability amplitudes through the factors $p_i^{q-1}$ and $p_i^{q-2}$.

For $q>1$, the exponent $q-1$ is positive, implying that low-probability amplitudes are strongly suppressed,
\begin{equation}
p_i^{q-1}\rightarrow 0
\qquad
(p_i\ll 1).
\end{equation}
Consequently, weak-amplitude fluctuations contribute only minimally to the entropy response. The entropy, therefore, becomes progressively less sensitive to perturbations as $q$ increases.

By contrast, for $q<1$, the exponent becomes negative,
\begin{equation}
p_i^{q-1}
=
\frac{1}{p_i^{1-q}},
\end{equation}
which amplifies the contribution of low-probability components. Small wavefunction amplitudes therefore produce disproportionately large entropy fluctuations, leading to enhanced sensitivity and numerical instability.

The entropic index $q$ thus acts as a fluctuation-resolution parameter: larger values of $q$ emphasize dominant wavefunction components, whereas smaller values increasingly probe weak-amplitude regions of the spectrum.

\subsection{Upper bound for entropy fluctuations}

The perturbative expansion also allows an explicit upper bound for the entropy variation to be derived. Retaining the leading-order contribution from Eq.~(\ref{eq:dSq}), one obtains
\begin{equation}
|\delta S_q|
\le
\frac{q}{|q-1|}
\sum_i p_i^{q-1} |\delta p_i|.
\label{eq:bound1}
\end{equation}

Applying the Cauchy-Schwarz inequality gives
\begin{equation}
\sum_i p_i^{q-1} |\delta p_i|
\le
\left(
\sum_i p_i^{2(q-1)}
\right)^{1/2}
\left(
\sum_i (\delta p_i)^2
\right)^{1/2}.
\end{equation}

Defining the perturbation norm
\begin{equation}
\|\delta p\|_2
=
\left(
\sum_i (\delta p_i)^2
\right)^{1/2},
\end{equation}
The entropy fluctuation satisfies the bound
\begin{equation}
|\delta S_q|
\le
\frac{q}{|q-1|}
\left(
\sum_i p_i^{2(q-1)}
\right)^{1/2}
\|\delta p\|_2.
\label{eq:mainbound}
\end{equation}

Equation~(\ref{eq:mainbound}) constitutes the central stability result of the entropy framework. The fluctuation amplitude is controlled by the spectral weight factor
\begin{equation}
\mathcal{F}_q
=
\sum_i p_i^{2(q-1)}.
\label{eq:Fq}
\end{equation}

For $q>1$, small probability amplitudes are increasingly suppressed inside $\mathcal{F}_q$, causing the fluctuation bound to decrease systematically with increasing $q$. Conversely, for $q<1$, weak amplitudes are amplified, and the fluctuation bound becomes increasingly large.

This analytical result directly explains the numerical behavior observed in Fig.~\ref{fig:fig1}(a), where the entropy deviation decreases systematically throughout the regime $q>1$.

\subsection{Thermodynamic scaling of the fluctuation bound}

One can further examine the scaling behavior of Eq.~(\ref{eq:mainbound}) in the thermodynamic limit.

In the extended regime, the probability distribution is approximately uniform,
\begin{equation}
p_i \sim \frac{1}{N},
\end{equation}
where $N$ denotes the system size. Substituting this form into Eq.~(\ref{eq:Fq}) gives
\begin{align}
\mathcal{F}_q
&\sim
N\left(\frac{1}{N}\right)^{2(q-1)}
\nonumber\\
&=
N^{3-2q}.
\end{align}

The fluctuation bound, therefore, scales as
\begin{equation}
|\delta S_q|
\lesssim
\frac{q}{|q-1|}
N^{\frac{3-2q}{2}}
\|\delta p\|_2.
\label{eq:scalingbound}
\end{equation}

The scaling form obtained in Eq.~(\ref{eq:scalingbound}) also allows a natural definition of an effective fluctuation scaling factor,
\begin{equation}
\Gamma_q(N)
\equiv
N^{\frac{3-2q}{2}},
\label{eq:Gammaq}
\end{equation}
such that the entropy fluctuation bound may be written compactly as
\begin{equation}
|\delta S_q|
\lesssim
\frac{q}{|q-1|}
\Gamma_q(N)\,
\|\delta p\|_2.
\label{eq:GammaBound}
\end{equation}

The quantity $\Gamma_q(N)$ directly characterizes the amplification or suppression of weak-amplitude fluctuations in the generalized entropy framework. Its scaling behavior reveals three distinct fluctuation sectors:

\begin{enumerate}
\item For $q<1$,
\begin{equation}
\Gamma_q(N)\sim N^{(3-2q)/2},
\end{equation}
grows rapidly with system size, implying strong amplification of weak-probability fluctuations.

\item For $1<q<3/2$, $\Gamma_q(N)$ remains algebraically increasing but with a reduced scaling exponent, corresponding to a fluctuation-suppressed yet finite-response regime.

\item At the critical threshold
\begin{equation}
q_c=\frac{3}{2},
\end{equation}
the scaling factor becomes system-size independent,
\begin{equation}
\Gamma_{q_c}(N)\sim \mathcal{O}(1).
\end{equation}

\item For $q>3/2$,
\begin{equation}
\Gamma_q(N)\sim N^{-(2q-3)/2},
\end{equation}
decreases with increasing system size, demonstrating asymptotic suppression of entropy fluctuations in the thermodynamic limit.
\end{enumerate}

The deformation parameter $q$ therefore acts as a fluctuation-control parameter governing the stability of generalized entropy diagnostics. Increasing $q$ progressively suppresses the contribution of weak-amplitude spectral components, thereby enhancing the robustness of entropy-based probes of localization and mobility-edge physics.

\subsection{Implications for mobility-edge detection}

The fluctuation analysis developed above has direct consequences for the entropy-gradient susceptibility introduced in Eq.~\ref{susceptibility}. The identification of mobility edges relies on resolving genuine energy-dependent restructuring of eigenstates across the spectrum. Such detection requires a stable entropy functional that is insensitive to the fluctuations arising from weak-amplitude components.

The analytical results derived above show that the regime $q>1$ naturally suppresses low-probability noise while retaining sensitivity to dominant wavefunction structure. Consequently, the entropy-gradient susceptibility exhibits sharp, stable signatures associated with spectral coexistence.

By contrast, in the regime $q<1$, amplified fluctuations originating from weak-amplitude components broaden the susceptibility response and affect the underlying mobility-edge structure.

The physically relevant regime for entropy-based mobility-edge detection, therefore, emerges naturally from the fluctuation stability properties of the generalized entropy itself. The entropy-gradient susceptibility is thus most effective precisely in the stability sector where perturbative entropy fluctuations remain intrinsically suppressed. The fluctuation analysis developed above also provides a broader perspective on the role of generalized entropy measures in spectral diagnostics. Different choices of the entropic index $q$ correspond to distinct amplitude sensitivities and therefore interpolate between different entropy frameworks. In the limit $q\to1$, the Tsallis entropy reduces to the Boltzmann-Gibbs-Shannon entropy, while generalized entropy measures such as the R\'enyi entropy (see Eq.~\ref{Re}) are similarly governed by the moments $\sum_i p_i^q$. The present analysis, therefore, establishes a stability-based criterion for identifying physically reliable entropy probes. In particular, the regime $q>1$ suppresses weak-amplitude fluctuations and yields robust spectral characterization, whereas the regime $q<1$ amplifies low-probability noise and produces unstable entropy responses. From this perspective, the fluctuation-stability framework serves as a natural selection criterion for generalized-entropy diagnostics in quasiperiodic systems. \begin{equation}\label{Re}
S_q^{(R)}
=
\frac{1}{1-q}
\ln\left(
\sum_i p_i^q
\right),
\end{equation}
Since the R\'enyi entropy depends on the same generalized moments
$\sum_i p_i^q$ appearing in the Tsallis entropy, its fluctuation
properties are governed by the same amplitude-weighting mechanism.
Consequently, the suppression of weak-amplitude fluctuations for $q>1$ and the enhancement for $q<1$ remain qualitatively unchanged for generalized entropy families constructed from identical probability moments. The principal numerical difference lies in the functional dependence on these moments: the Rényi entropy applies a logarithmic transformation,
while the Tsallis entropy is linear in the same generalized probability moment. As a consequence, the logarithmic mapping compresses the numerical response to perturbations, while the linear form preserves perturbative variations in a more direct algebraic form. This makes the Tsallis entropy particularly convenient for deriving the analytical fluctuation bounds developed in the present work and
for quantifying spectral restructuring via the entropy gradient susceptibility, while the underlying amplitude-weighting mechanism responsible for the fluctuation-stability crossover is common to both entropy families, their quantitative fluctuation responses differ due to their distinct functional dependence on the generalized probability moments.
\section{Numerical Results}
\label{sec_4}

We now investigate the fluctuation-stable entropy framework in quasiperiodic systems exhibiting qualitatively distinct localization behavior: the Aubry-Andr\'e (AA) model with a homogeneous localization transition and the generalized Aubry-Andr\'e (GAA) model hosting mobility edges. To quantify fluctuation stability, weak random perturbations of amplitude $\epsilon=10^{-5}$ were added to the probability distribution of each eigenstate, followed by renormalization to preserve $\sum_i p_i=1$. The entropy deviation $\langle |\Delta \tilde S_q| \rangle$ was then obtained by averaging over all eigenstates and six uniformly distributed quasiperiodic phases $\phi\in[0,2\pi)$ for a system size $L=600$.

Figure~\ref{fig:fig1}(a) first illustrates the fluctuation properties of the normalized Tsallis entropy under weak perturbations of the wavefunction amplitudes. The average entropy deviation $\langle |\Delta \tilde S_q| \rangle$ decreases systematically throughout the regime $q>1$, demonstrating that generalized entropy fluctuations are progressively suppressed as the entropic index increases. By contrast, the regime $q<1$ exhibits enhanced sensitivity due to the amplification of low-probability wavefunction components. These results directly confirm the analytical stability framework derived in Sec.~\ref{sec_3} and identify the regime $q>1$ as the physically stable sector for entropy-based spectral probes.

The consequences of this stability crossover become evident in the entropy-gradient susceptibility defined in Eq.~\ref{susceptibility}, which probes the energy-dependent restructuring of eigenstates across the spectrum. Figure~\ref{fig:fig1}(b) shows that the susceptibility profile becomes increasingly sharp and stable for $q>1$, whereas the regime $q<1$ develops broad responses. The suppression of weak-amplitude noise in the stable entropic sector, therefore, yields enhanced spectral resolution.

The distinction between homogeneous localization and mobility-edge physics is summarized in Fig.~\ref{fig:fig3}. In the AA model, shown in Fig.~\ref{fig:fig3}(a), the susceptibility exhibits only a smooth crossover near the global transition point $\lambda_c=2t$, reflecting the absence of energy-dependent coexistence within the spectrum. The normalized entropy profile shown in Fig.~\ref{fig:fig2}(a) correspondingly evolves smoothly across the transition without developing pronounced spectral structure.

By contrast, the GAA model develops a sharp and systematically size-stable susceptibility peak within the coexistence regime, as shown in Fig.~\ref{fig:fig3}(b). This behavior originates from the rapid restructuring of localized and extended eigenstates across the spectrum near the mobility edge. In the inset, we also show the IPR variance plot, where, for the AA model, we observe a jump in the variance from 0 to a finite value at the transition point $\lambda_c = 2t$. However, in GAA, there is no characteristic feature that can indicate the presence of a mobility edge. The contour plot of $\tilde S_q(E,\lambda)$ in Fig.~\ref{fig:fig2}(b) further reveals a clear energy-dependent separation between localized and extended regions, consistent with the analytical mobility-edge condition
\begin{equation}
\alpha E = 2t-\lambda,
\end{equation}
represented in a blue dashed line over the plot. $\lambda_{c_1}$ and $\lambda_{c_2}$ are the critical strengths between which the mobility edge exists in the spectrum, and the susceptibility peak persists. We see a similar behavior in an SSH model with a quasi-periodic onsite potential, which is known to exhibit a mobility edge. The numerical results, therefore, demonstrate that fluctuation suppression and spectral sensitivity emerge together within the stable entropy sector. While homogeneous localization transitions produce only weak susceptibility responses, mobility-edge systems generate pronounced and robust entropy-gradient signatures associated with spectral coexistence. 

\section{Analytical insights from limiting cases}
\label{sec_5}

The physical origin of the entropy-gradient susceptibility $\chi_q$ can be understood by considering three limiting spectral regimes: fully extended states, strongly localized states, and spectra containing mobility edges. These limits clarify why $\chi_q$ remains weak in spectrally homogeneous phases but develops pronounced peaks in the presence of spectral coexistence.

In the fully extended regime, the probability distribution is approximately uniform,
\begin{equation}
p_i \simeq \frac{1}{N},
\end{equation}
yielding the maximal Tsallis entropy,
\begin{equation}
S_q = S_q^{\mathrm{max}},
\qquad
\tilde S_q \simeq 1.
\end{equation}
Since the entropy profile remains nearly constant across the spectrum, its energy derivative is strongly suppressed,
\begin{equation}
\chi_q \rightarrow 0,
\qquad \text{(extended regime)}.
\end{equation}

In the strongly localized regime, eigenstates occupy an effective localization volume $\xi \ll N$, giving
\begin{equation}
\tilde S_q
\sim
\left(\frac{\xi}{N}\right)^{1-q}.
\end{equation}
Although the entropy becomes small, the localization length remains approximately uniform throughout the spectrum, so $\tilde S_q(E)$ again varies weakly with energy. Consequently, $\chi_q$ remains small in homogeneous localized phases.

The situation changes qualitatively in the presence of a mobility edge, where localized and extended states coexist within the same spectrum. Approximating the coarse-grained entropy profile as
\begin{equation}
\tilde S_q(E)
\simeq
f(E)S_E+[1-f(E)]S_L,
\end{equation}
where $f(E)$ denotes the fraction of extended states below energy $E$, differentiation gives
\begin{equation}
\frac{d\tilde S_q}{dE}
=
\frac{df}{dE}(S_E-S_L).
\end{equation}
If the crossover occurs over an energy scale $\Delta E$, then
\begin{equation}
\chi_q
\sim
\frac{|S_E-S_L|}{\Delta E}\,f(1-f).
\label{eq:chi_scaling}
\end{equation}

Equation~(\ref{eq:chi_scaling}) shows that the susceptibility is governed by three ingredients: the entropy contrast between localized and extended states, the coexistence factor $f(1-f)$, and the sharpness of the spectral crossover. The susceptibility is therefore maximized in the mobility-edge regime where both classes of states coexist in comparable proportions. By contrast, homogeneous spectra with either $f=0$ or $f=1$ suppress the coexistence factor, leading only to smooth crossover behavior.

These considerations demonstrate that $\chi_q$ directly probes spectral heterogeneity through the energy-dependent restructuring of eigenstates. Within the stable regime $q>1$, the resulting susceptibility peaks provide a robust signature of mobility-edge physics.

\section{Conclusion} \label{sec_6}

We developed a fluctuation-stability theory for generalized-entropy diagnostics and introduced the entropy-gradient susceptibility, $\chi_q$, as an energy-resolved probe of spectral heterogeneity in quasiperiodic systems. Our analysis shows that weak-amplitude fluctuations are naturally amplified for $q<1$ and suppressed for $q>1$, thereby providing a theoretical foundation for the physically robust $q>1$ sector widely employed in generalized participation and multifractal diagnostics. A thermodynamic scaling analysis further identified an asymptotically stable regime for $q>3/2$.

As a physical application of this framework, we demonstrated that the entropy-gradient susceptibility directly captures spectral restructuring across the quasiperiodic spectrum without requiring explicit classification of individual eigenstates. In the Aubry-Andr\'e model, where the localization transition is spectrally homogeneous, the susceptibility exhibits only weak crossover behavior. In contrast, the generalized Aubry-Andr\'e model exhibits sharp, systematically stable susceptibility peaks within the mobility-edge regime due to the coexistence of localized and extended states across different energy sectors.

Our results establish fluctuation stability as a general guiding principle for entropy-based diagnostics across disordered quantum systems, not limited to quasiperiodic models. A central outcome of this work is that the commonly employed \(q>1\) sector of generalized participation and multifractal diagnostics emerges naturally as the physically robust fluctuation-stable regime of the entropy framework. Since mobility edges have already been observed in cold-atom and photonic quasiperiodic lattices, the present framework also provides a direct route to the experimental investigation of fluctuation-stable entropy signatures in synthetic quantum systems. Future work may extend this framework to interacting systems, thereby further broadening its relevance to condensed-matter and quantum-simulation platforms.

\section{Acknowledgments}
AG gratefully acknowledges Dr. Shaon Sahoo for the discussions on the related topic. 
\bibliographystyle{plain}        % Include this if you use bibtex 
\bibliography{manuscript}

\end{document}